\documentclass[letterpaper,english,floatfix]{revtex4}
\usepackage{pslatex}

\makeatletter

\usepackage{bm}
\usepackage{babel}
\makeatother
\begin{document}

\title{Stochastic model Of universe which constantly creates dark
energy (Omega=0.7) and dark matter (Omega=0.3) but instantly at
0.12Gyr created nucleons and radiation.}

\author{\textbf{Z. Alexandrowicz}}

\email{bpalexan@weizmann.ac.il}

\affiliation{\textbf{Weizmann Institute of Science, Materials and
Interfaces, Rehovot 76100, Israel}}

\begin{abstract}


\leftskip 0.6in. \hsize 6.6in.

An assumption attributing vacuum mass energy to symmetric 'Null'
fluctuation which with equal probability either adds or subtracts,
virtual planck either particles or antiparticles leads to the
following net resultant: 'Dark Energy' virtual
particle-antiparticle pairs and 'Dark Matter' real planck
particles, constantly give $\Omega_{DE}=0.7$ and $\Omega_{DM}=03$.
Second assumption that gravitational attraction propagates as
particle-bridging wavelength, leads to following estimates:
Instantly, astonishingly recently and utilizing the wavelength,
the Dark Energy particles converted into nucleons(n) and the
antiparticles into radiation (ultimately CMB). Baryogenesis
occurred almost immediately in their ultra-hot clusters. Nucleonic
matter declined from $\Omega_n=0.7$ at creation, to present time
$\Omega_n=0.03$. Famous 'Acceleration' is attributed to the
cohesion of a supreme cluster of nucleonic matter, giving
$z/\ell=(1-\ell/\ell_H)^{2/3}$, $z,$ $\ell$ and $\ell_H$ being
respectively, redshift, luminosity and Hubble distances.

\end{abstract}

\maketitle

\leftskip 0.4in. \hsize 6.0in.





\noindent{\bf $1$ INTRODUCTION.}

The new Big Bang (BB) model leaves certain questions unanswered
\cite{book}. The Universe geometry seems Flat and the density of
matter mostly non-nucleonic Dark Matter (DM) is almost but not
quite critical at $\Omega_{dm}\simeq 0.3.$ The deficit is
attributed to Dark Energy (DE) which provides $\Omega_{de}\simeq
0.7.$ However the two $\Omega's$ should evolve entirely
differently in time so how can we explain their present
near-coincidence. Also what does $\Omega_{de}$ represent. When it
is attributed to vacuum energy fluctuation (VF) its energy density
seems to exceed today's $\Omega\simeq 1$ by $10^{120}.$ I propose
a new approach to these questions with the help of a stochastic
model of Universe expanding with constant speed. Deceleration due
to gravity is balanced by mass energy constantly created by VF
which constantly maintains $\Omega=1.$ (Permanent but different
creation has been proposed by Hoyle and
Narlikar\cite{hoyle},\cite{narlikar}). Formally the subject
studied here belongs to quantum gravity but in order to bring the
two together my model starts straightaway from the Planck
particle\cite{book},\cite{narlikar}. Often expressed opinion is
that VF had created just after BB and possibly continues to create
virtual planck particles and antiparticles $\bar p$ and $\tilde p$
respectively. These decay after the planck time and Dark Energy
density is simply equated to that of a planck particle. I claim
instead, that VF constitutes a four sided Null vacuum fluctuation
(NVF) which with an equal probability generates at all times
$+\bar p,$ or $-\bar p,$ or $+\tilde p,$ or $-\tilde p.$ In this
context for example $-\bar p$ does not signify decay of an
existing $+\bar p,$ but merely a decrease by one of the total
number of $\bar p.$ Indeed it may be followed by another $-\bar p$
on the same site. To first approximation the sum of the
fluctuations is null but their so called root mean square $(rms)$
resultant does not vanish. It is to it that I attribute constant
creation of mass energy, DM and DE alike.

To begin with let us assume a constantly Flat expanding universe
and express our results in terms of a reduced dimensionless
radius-time $r,$ which constitutes a ratio of the causal
('Hubble') radius and the planck radius $r_p.$ Henceforth all
distances, times and masses are reduced by corresponding planck
quantities. Total number of all planck fluctuations ($\pm\bar p$
and $\pm\tilde p$) created in the sphere demarcated by the causal
radius and during epoch $r,$ is $r^4.$ However the $rms$ number of
excess $+\bar p$ and $+\tilde p$ fluctuations referred to jointly
as excess $p_+$ particles, is only $r^2/2.$ They or rather signals
they have emitted over epoch $r$ give rise to cohesive energy
$-\Gamma=G\sum m_i/r_i$ acting on an arbitrary 'central particle'.
Here $m_i$ is the mass of an excess $p_+$ in causal contact with a
central particle over a distance $r_i$ during unit present time.
Excess $p_+$ play a double role, first as {\it past} particles
which via signals contribute to $\Gamma$ and second as {\it
present time} central particles acted upon by $\Gamma.$ Due to
their unit duration the number of excess $p_+$'s in causal contact
with a center during unit time is $r/2$ instead of $r^2/2.$ We do
find that $-\Gamma=G\sum m_i/r_i=G/2,$ get a permanently critical
energy density\cite{book}), in agreement with our starting
assumption of constantly Flat universe.

I assume further that actually  the excess $p_+$ particles consist
of two fractions: The one consists of virtual $(\bar p-\tilde
p)_+$ pairs and corresponds to DE. The other fraction consists of
real planck particles $\bar p_+$ and corresponds to DM. The
fractions' relative abundance $0.69:0.31$ is estimated with the
help of a simple Lattice Model of randomly distributed paired and
single, particles and antiparticles. Despite profound difference
between short lived virtual $(\bar p-\tilde p)_+$ pairs and the
permanent real  $\bar p_+$'s,  both play the same role in
expanding Universe: their uniform creation keeps $\Gamma$ constant
and the expansion 'balanced', neither decelerated nor accelerated.
Notwithstanding the virtual versus real disparity their effective
contribution to $\sum m_i/r_i$ is still in the $0.69:0.31$ ratio.
Thus the number of the real (DM) $\bar p_+$'s created is $r$ times
smaller than that of the virtual (DE) $(\bar p-\tilde p)_+$ but is
compensated by their Action which continues indefinitely during
$r.$ The overall picture is of a featureless balanced universe
which expands without big bang or big inflation. Yet the presence
of cosmic microwave background CMB and of nucleonic matter $r_i$
remains unexplained. Their creation is studied here in detail as
an important example of momentary 'transitions' viz., deviations
from the uniform creation of DE and DM.

In order to explain a transition the basic assumption is amplified
by assumptions on the nature of signals by means of which $p_+$'s
which existed in the past, attract $p_+$'s existing at present
time. Since quite a few assumptions are introduced before complete
picture starts to emerge let me preview the argument briefly. The
massless signals propagate indefinitely in spacetime with speed
$c$ and may be associated with a wavelength $\lambda.$ Avoiding
reference to a 'graviton' I only assume that in order to propagate
the massless signals have to act on masses uninterruptedly viz.,
on 'central particles {\it spaced at intervals equal to the
$\lambda$ wavelength}. The process creates the cohesive energy
$-\Gamma.$ It soon becomes clear that the $r$ excess $p_+$ spread
over the $r^3$ causal sphere are too sparsely spaced to meet this
requirement. I assume that the role of past-attracting-present
particles is played by virtual particles denoted $q,$ generated
from excess $p_+.$ It transpires that the radius-lifetime of a $q$
particle has to be equal to $s,$ where $s^2$ is equal to the
separation distance between the excess $p_+$ viz.,
$s^2=(r^3/r)^{1/3}.$ In that case we can show that everywhere in
space $\lambda$-$q$ couples are created one after another along
signals' straight line path. However the argument requires several
more steps. I assume that each excess $p_+$ is generated by NVF in
a time sequence of interconnected steps, creating on the average
one $q$ per step. The $s$ longer than planck's radius-lifetime of
a $q$ particle implies by Uncertainty an $s$ times smaller mass
viz., $m_q=m_p/s.$ Hence $s$ virtual $q$'s may be generated out of
each excess $p_+$ at distance $s=\lambda$ from each other. A
question arises how are the $q$ particles generated precisely
along the straight path of $\lambda$'s. Since the two jointly
create an increment of cohesive energy I assume that such local
minimum favors the creation of $q$'s one after another along the
straight path of $\lambda$'s. In return the creation of $q$'s
enables the propagation of massless $\lambda$'s from one mass to
another. This straight low energy path of sequentially created
$\lambda-q$ couples is named here a 'channel'. However the
channels cannot propagate in isolated straight lines over
radius-time $r;$ they have to create a contiguously connected
network in the $r^3$ space, without big holes. Comparison to
percolating clusters indicates that the stepwise generation of
$q$'s which {\it on the average} creates one $q$ per step, has to
bifurcate repeatedly creating a space filling (fractal) Cluster of
interconnected divergent channels. The clusters have to be large
enough in order to also be contiguously interconnected. The
restrictions put together enable one to estimate the number and
radius-lifetime of clusters and of their $s$ constituent
$\lambda-q$ couples.

The results enable me to propose a model of a nucleon and CMB
creating transition.  Major role belongs to scales whose present
time values are: radius $s\simeq10^{20}$ of a $\lambda-q$ couple;
radius $s^{3/2}\simeq10^{30}$ of a cluster and distance
$s^2\simeq10^{40}$ between neighbor excess $p_+$. The numbers and
their combinations bring to mind the large dimensionless numbers
to which Dirac\cite{dirac} has attributed major cosmic
significance. His numbers are on the order of $10^{40}$ and
represent the ratio of planck to proton masses, of cosmic to
proton radii, etc. The hint worked out systematically led me to
the conclusion that not today but quite recently when the universe
was merely one hundred times younger than now and $s$ was equal to
$0.4\times 10^{20}$ $(10^{-13}cm),$ the energy generated by the
$\lambda$ to $q$ coupling became precisely equal to the rest mass
of a nucleon. At that moment which lasted $14$ seconds occurred a
transition which in each cluster diverted the generation of DE,
into a generation of nucleons and of high energy radiation (which
cascaded down to CMB). Simultaneously in same clusters occurred
Baryogenesis. Such description of baryogenesis is very different
from standard BBN theory but on the face of it seems to fit known
facts. All these results are obtained with the sole input of
proton's mass. Injection of observed present time value of $r_0$
allows us to estimate today's $T_0$ and the dilution of nucleonic
matter in expanding space. We find the recently reported
Acceleration of expanding Universe to be illusory and attributable
to self-attraction of nucleons which binds them together retarding
dilution.

\noindent{\bf $2$ UNIFORM GENERATION OF MASS AND ENERGY BY NVF.}

  Our story unfolds as follows. The effect of gravity
on the expansion is represented by a cohesive negative energy
$m_p\Gamma\equiv-m_pG\sum_i m_i/r_i.$ It results from an
attraction of an arbitrary central planck mass $m_p$ during unit
present time by surrounding masses $m_i$ acting over corresponding
distances $r_i.$  Here $m_i$ and $r_i$ are reduced by
corresponding planck quantities however the single mass $m_p$ of
the central particle is not reduced in order to display its
presence. We sort $\sum_i m_i/r_i$ into concentric shells around
the center, $1\le r'<r.$ An $r'$'th shell contains all particles
virtual and real which existed at $r'$'th past time and are in
causal contact of unit duration with the central $m_p$ at a
present time $r,$ over distance $r-r'.$ At this stage we avoid
explicit mention of causal signals. Also in order to simplify the
verbal argument all particles are attributed a unit lifetime and
those 'existing at..' are identified with 'created' at..'.
Denoting the contribution of an $r'$'th shell by $g_{r'},$ we
reexpress the cohesive energy 'per (central) mass' (abbreviated to
$'pm$-energy') as $-\Gamma=G\sum m_i/r_i=G\int g_{r'} dr'.$ The
planck particle mass, radius and lifetime used here namely $m_p,$
$r_p$ and $t_p=r_p/c$ respectively, are defined by identifying a
cohesive $pm$-energy created by one planck particle with its rest
energy namely $Gm_p/r_p=c^2$ and by the Uncertainty relationship
$Gm_p^2=\hbar c.$ In view of $Gm_p/r_p=c^2,$ non-reduced
$pm$-energy becomes $-\Gamma\equiv c^2\int g_{r'}dr'.$ To begin
with we focus on planck particles constantly generated by VF
aiming to show that $\int g_{r'}dr'=1/2,$ a value which keeps
$\Gamma$ constant in expanding universe.

A $\Gamma=const$ result has been derived with the help of
dimensional analysis by Chen and Wu\cite{chen-wu} extended by John
and Joseph\cite{john-joseph}, but here the stochastic nature of VF
is revised for that purpose. It is commonly held that VF
constitutes a birth and death process: Creation of virtual $\bar
p\tilde p$ pairs with constant probability density, with each
creation followed within $t_p$ by inevitable extinction. Hence VF
is believed to create an energy density equal to that of a single
planck particle, a stunning overestimate of today's
value\cite{book}. I propose to overcome this problem by redefining
VF as Null vacuum fluctuation NVF which with an equal probability
density, adds or subtracts either $\bar p$ or $\tilde p.$  To
first approximation it gives null macroscopic variation of mass
energy. Equal probability however does not imply neither that $+$
and $-$ occur in precisely equal numbers nor that $\bar p$ and
$\tilde p$ do. It leaves room for deviations estimated as $rms$
(root mean square) resultant. Thus NVF constitutes a {\it critical
fluctuation} like in collective phenomena. Although perhaps
unorthodox this seems to fit the isolated universe.

In order to  show that $\int g_{r'}dr'=1/2$ we temporarily ignore
the distinction between particles and antiparticles, lump together
the $\pm\bar p$ and $\pm\tilde p$ fluctuations, use a $p_{\pm}$
common notation and concentrate only on a deviation from the $\pm$
symmetry. Random Walk provides an example for calculating the
$rms$ resultant. Let $N$ be the number of random back and forth
$b_{\pm}$ steps. Due to mutual cancellation their $rms$ resultant
denoted here as $\langle b_{\pm}N\rangle$ is equal to $b_+
N^{1/2}$ only. We return to  NVF. The number of  $p_{\pm}$
fluctuations of unit duration in a hyper-sphere of radius $r$ is
$r^4$ but the $rms$ number of excess $p_+$ at radius time $r$ is
only $r^2/2$ or in detail, $r'$ particles created at past time
$r'$ and integrated over $1\le r'<r.$ The $r'$ have been created
randomly distributed over $r'$ distinct distances from the center.
Only one manages to establish causal contact with the center at
time $r$ over distance-time $(r-r')$ with probability  $1/r'.$ The
$pm$-energy associated with this lucky increment is proportional
to $(r-r')^{-1}.$ Assuming uniform expansion (justified a
posteriori), the latter is equal to $r'/r.$ Thus the contribution
of $r'$ particles created at time $r'$ to $pm$-energy becomes
$(r'/r')(r'/r)$ giving $g_{r'}=r'/r.$ The lifetime of the $p_+$
particles hence the attraction they exert on each other lasts one
unit of time, $t_p$ (using for the moment unreduced time). Hence
the increment of $pm$-Action created by each planck mass is
$(Gm_p/r_p)t_p=c^2t_p.$ This increment is assumedly conserved
since uniform expansion performs no work (once more justified a
posteriori). However the range and epoch of the Action of past
$p_+$ on present ones increases without bounds with $r.$
Uncertainty requires $\bar h=(m_p c^2)(t_p).$ As time of Action is
dilated to $rt_p,$ the $pm$-energy decreases to $m_pc^2/r.$ Hence
in order to sum the $pm$-energy of all $1\le r'<r$ past particles
we integrate over a fraction-time $r'/r.$ We refer to this
decrease as 'range-decimation'. We get $\int
g_{r'}dr'=\int(r'/r)d(r'/r)=1/2$ and $-\Gamma=c^2/2=Gm_p/2r_p.$ We
also wish to evaluate an 'objective' mass energy of excess $p_+$'s
irrespective of their irrespective of their contribution to
$-\Gamma.$ to a particular center. To this end we have to factor
out from $g_{r'},$ the aforementioned $1/r'$ and $r'/r$ factors.
We get $g_{r',mass}=r'.$ Integrating over $r'/r$ we get total
reduced mass $M$ of excess $p_+$ contributing to $\Gamma,$ namely
$M=r/2$ just the hoped-for result. Summarizing with the help of
random walk symbols we write
\begin{equation}
\label{gamma} -\Gamma/c^2=\langle r^4 p_{\pm}\rangle=
(\int_{1/r}^{1-1/r}{~r'\over r}d{~r'\over r})p_+={1\over 2}p_+
~~or~~-\Gamma={Gm_p\over 2r_p}={c^2\over 2}~~and~~M={r\over 2}.
\end{equation}
The 'range-decimation' plays a major role in Eq.(\ref{gamma}): it
turns an open ended integral over $1\le r'<r$ unit times into a
bounded $r$ times smaller integral inside $1\le r'/r\le 1-1/r.$ It
applies to the case of increments of $pm$-Action created by past
virtual particles, whose initial time duration was only one $t_p,$
but whose past-present $pm$-Action increases indefinitely with
$rt_p.$ However {\it if the creation of $pm$-Action by a past
particle coincides with the range of past-present contact, the
associated $pm$-energy is not range-decimated} (a case arising
with a real past particle).

For certain purposes we are not interested in the invariant
balance. We are interested instead in properties which in effect
belong entirely to a present time viz., to the upper limit in
Eq.(\ref{gamma}). For example particles' {\it rest mass}, number
of planck masses created by NVF at anytime, their energy density,
temperature, mutual interaction etc. In that case the integration
of Eq.(\ref{gamma}) becomes redundant. Renaming its upper limit as
'now' I propose that the properties just listed are determined by
$\Gamma_{now}$ and $M_{now},$ defined as follows
\begin{equation}\label{mach}
-\Gamma_{now}/c^2=r^{-1}d(\smallint^{now}r'dr')/d(now)=1;
~~(m)c^2=-(m)\Gamma_{now}~~and~~M_{now}/r=2M/r=1.
\end{equation}
Verbally, rest energy of mass $m$ is equal to $m$ multiplied by
$pm$-energy $-\Gamma_{now}.$ The product expresses their present
time mutual interaction as opposed to the description of balanced
expansion during epoch $r.$ The  $pm$-energy $\Gamma_{now}$ is
twice as large as $\Gamma.$ It implies a modified Mach principle
saying: "Rest energy $mc^2$ is due to a limiting present time
attraction of mass $m$ by the NVF-created $pm$ cohesive energy
$-\Gamma_{now}=Gm_p/r_p.$ " Incidentally $\Gamma_{now}$ also
obtains at an $r=1$ discrete limit of the causal radius, viz.,
there is no singularity if we believe that space must generate
planck fluctuations.

\noindent{\bf $3$ DARK ENERGY AND DARK MATTER; CRITICAL ENERGY
DENSITY.}

The $rms$ resultant of Eq.(\ref{gamma}) lumps together $\bar p$
and $\tilde p$ excess fluctuations into $p$ ones. In order to
progress further we have to sort this resultant into two
fractions. A $\theta_1$ fraction results from asynchronous $\pm$
fluctuation of single $\bar p$ and in separate, of single $\tilde
p$ and leads to an excess of  $\bar p_+$ and in separate of
(debit) $\tilde p_-,$ denoted $\bar p_+\Vert\tilde p_-$. A
$\theta_2$ fraction results from synchronous joint $++$ or joint
$--$ fluctuation of paired neighbor $\bar p$ and $\tilde p$ and
leads to an excess of paired $+\bar p$ and $+\tilde p,$ denoted
$(\bar p-\tilde p)_+$. We utilize a cubic Lattice Model whose
space filling cubic sites are occupied each by either $\bar p$ or
$\tilde p$ with equal probability (signs undetermined). A priori
each of the $\bar p$(or $\tilde p)$ is 'single' with probability
$\theta_1$ or, it is 'paired' to an 'anti' neighbor with
$\theta_2(=1-\theta_1).$ We estimate $\theta_1$ with the help of a
Gedanken simulation. We pick at random a $\bar p_{old}$ and flip
it over to $\tilde p_{new}$ (or vice versa). Before the flip $\bar
p_{old}$ was either 'single' or 'paired'. In the first case
$\tilde p_{new}$ creates a pair if at least one of its six
neighbors corresponds to a single $\bar p.$ The respective
'mean-field' probabilities that one and that all single neighbors
are $\tilde p$  is $0.5$ and $0.5^{6\theta_1}$ respectively. Thus
$t_{1,2}=1-0.5^{6\theta_1}$ is the transition probability that
$\tilde p_{new}$ and its single neighbor which is $\bar p$ are
declared 'paired'. In the second case $t_{2,1}=0.5^{5\theta_1}$ is
the transition probability that a $\tilde p_{new}$ which before
the flip constituted a paired $\bar p_{old},$ will not pair with
any of its single neighbors because they all are $\tilde p$ (the
old partner is so automatically), whereupon $\tilde p_{new}$ and
its old partner are declared 'single'. Else there occurs a mere
exchange of partners. Detailed balance is
\begin{equation}
\label{theta} \theta_2/\theta_1=t_{1,2}/t_{2,1}~~giving~~\theta_2
\simeq 0.685~~(\theta_1+\theta_2\equiv 1).
\end{equation}
The synchronous evolution of the $\theta_2$ fraction breaks the
$\pm$ symmetry but conserves the particle antiparticle one. Its
$rms$ resultant consists of virtual particles and their anti-
linked together in $(\bar p-\tilde p)_+$ pairs. Hence their
contribution to $\Gamma$ is identical to that in Eq.(\ref{gamma}),
except that here it is limited to a $\theta_2$ fraction of all
fluctuations. The momentary lifetime of $(\bar p-\tilde p)_+$
pairs precludes their correlation to cosmic structure. Hence we
identify the $\theta_2$ fraction with DE. Using previous symbols
we write
\begin{equation}
\label{de}-\theta_2\Gamma/c^2=\big\langle(\bar p_{\pm}+\tilde
p_{\pm})r^4\big\rangle_{syn}=(1/2)(\bar p-\tilde
p)_{+}~~giving~~M_{de}/r=\theta_2/2.
\end{equation}
The asynchronous evolution of the $\theta_1$ fraction is divided
into hypothetical stages. The $rms$ resultant of first stage which
has been denoted $\bar p_+\Vert\tilde p_-,$ breaks the particle
and anti- symmetry and consists of excess $\bar p_+$ but also with
an equal probability, of excess $\tilde p_-$ 'debit' fluctuation.
Interpreting the resultant as net gain of $\bar p_+$'s combined
with an equally probable net loss of $m_p$ masses whatever be
their source, we view the combination as equivalent to $\bar
p_{\pm}.$ The number of these variations is $\theta_1 r^2/2.$ We
discard alternative resultants consisting either of $\bar
p_-\Vert\tilde p_+$ or of $\bar p_+\Vert\tilde p_+$ which would
self-annihilate in our particle dominated universe letting NVF
generate vast positive energy gratuitously. Our intermediate $\bar
p_{\pm}$ resultant has an approximately null rest mass and exerts
null attraction. In the second stage the average $(r^2/2)\bar
p_{\pm}$ produces an $rms$ $(r/2)\bar p_+$ net positive but $r$
times smaller mass. (The one half factor is not affected because
one of the two averages  should involve $M_{now}$). In the absence
of minus variations and of antiparticles the resultant consists of
real particles of planck mass. As we have pointed out after
Eq.(\ref{gamma}) the contribution of a real particle to
$pm$-energy $\Gamma$ lasts indefinitely hence in contrast to a
virtual particle, is not subject to 'range-decimation'.
Alternatively it is convenient to view each real particle as
equivalent to a sequence of $r$ 'Xerox copies' reproducing its
mass $r$ times over again one atop another, each copy of unit
duration. In that case the second averaging does not decrease the
number of particles by $r$ but range-decimation does so instead.
One way or another a conversion of $r^2/2$ virtual into $r/2$ real
$\bar p$ is costless, yielding an $r/2$ and an $1/2$ result for
$M_{dm}$ and $\Gamma$ respectively. The extraordinary mass and
presumably inertness of real planck particles suggests that they
be identified with DM, which participates in cosmic structures.
Using previous symbols we write
\begin{equation}
\label{dm} -\theta_1\Gamma/c^2=\big\langle(\bar p_{\pm}+\tilde
p_{\pm}) r^4\big\rangle_{asyn}=\big\langle\bar
p_{\pm}r^2\big\rangle/2=(1/2)\bar
p_{real};~~~~M_{dm}/r=\theta_1/2.
\end{equation}
Jointly $M_{de}/r$ and $M_{dm}/r$ give $M/r=1/2$ and a
$\rho_p=(m_p r/2)(4\pi/3)^{-1}(rr_p)^{-3}$ joint mass energy
density. Using $m_p/r_p=c^2/G,~$ $c=r_p/t_p,$ $(rt_p)^{-1}\equiv
H$ (the Hubble coefficient) and the common notation
$\Omega\equiv\rho/\rho_c,$ we find our $\rho_p$ equal to
conventional critical density $\rho_c\equiv 3H^2/8\pi G.$ Thus
\begin{equation}
\label{omega} \rho_p=(m_p/r^2)(3/(8\pi
r_p^3)=\rho_c(always)~~~or~~~ \Omega_p=\Omega_{de}+\Omega_{dm}=1.
\end{equation}
Here $\Omega_{de}=\theta_2,$ $\Omega_{dm}=\theta_1$ of
Eq.(\ref{theta}). Interestingly critical density $\rho_c$ holds at
all times and  is simply equal to the density of planck particles
$\rho_p$ constantly created by NVF. Constantly critical $\rho_p$
implies constant validity of the Friedmann equation describing
Flat universe. This is consistent with our starting assumption of
constant Flatness and uniform expansion. It should be stressed
that here Dark Energy and Dark Matter constitute two facets of the
mass energy constantly created by NQF whose {\it joint uniform
generation} keeps $M/r$ hence $\Gamma$ constant. This contrasts
the role usually attributed to DE namely that of a
Universe-Expanding and Matter-Opposing dark energy. In view of our
results a more fitting appellation doing justice to the common
role and origin of DE and DM would respectively be Virtual and
Real excess planck particles  constantly created by NVF.

\noindent{\bf $4$  $\lambda$'s AND $q$'s ON SCALE $r^{1/3},$ IN
$\kappa$-CLUSTERS ON SCALE $r^{1/2}.$}

Once more in order to progress further we have to elaborate a
previously simplified description. We still focus on the simpler
case of DE consisting of virtual paired planck particles and
antiparticles denoted jointly as $p_+$ like in Eq.(\ref{gamma}).
Later it will transpire that our conclusions to DM as well. Thus
each $p_+$ has the planck unit rest mass, unit lifetime and prior
to extinction emits a signal which inherits the $c^2(=Gm_p/r_p)$
unit cohesive $pm$-energy of the parental $m_p$ mass. During
emission $pm$-energy and $pm$-Action coincide numerically. The
following properties of signals have been attributed before to
anonymous 'causal connection' between past and present planck
particles. The signals are massless, propagate with speed $c$ a
$pm$-energy and in order to create energy itself have to interact
with some intercepting particles. Are these indeed the planck
particles? The two defining equations $Gm_p/r_p=c^2$ and
$Gm_p^2=\hbar c$ single out the planck $m_p$ and $r_p$ as the
values fitting an emission of signals having the $c^2$
$pm$-energy, which to recall endows any intercepting mass $m$ with
the $mc^2$ rest energy (Eq.\ref{mach} and the modified Mach
principle). Therefore in Eq.(\ref{gamma}) we have attributed to
excess $p_+$'s both the emission and interception of signals.
However the $p_+$'s are too short-lived for that. I therefore
assume that to begin with NVF generates as its $rms$ a longer
living particle (denoted $q)$ which performs the role of signal
emission and interception. The $q$-particles are generated in a
{\it cluster} denoted $\kappa,$ such that total mass of each
$\kappa$-cluster adds up to precisely one $m_p.$ The number, size,
structure and generation of the $\kappa$-cluster have to meet
certain requirements based on the following considerations.

  {\it $\lambda$ wavelength and $q$ particle both
on scale $s=r^{1/3}$.} The signals may be associated with a
wavelength $\lambda$ describing  one cycle variation of their
$c^2$ $pm$-energy. Equations (\ref{gamma}) and (\ref{mach}) show
that the dominant contribution to cohesive $pm$-energy is due to
signals which propagate over a relatively short past-present time
separation $r'/r\rightarrow 1).$ Let us look at the distance-time
separating present time neighbor $p_+$'s. Their number is $r$
(Eq.\ref{mach}). The corresponding inter-$p_+$ distance is
$R_{pp}=(r^3/r)^{1/3}=r^{2/3}.$  We denote $r^{2/3}=s^2$ ($\sim
100km$ today). Identification of the $R_{pp}$ distance with
$\lambda$ is ruled out. It implies that rest mass of matter may
fluctuate below an incredibly large limit. I conjecture that both
$\lambda$ and the radius-lifetime $r_q$ of the $q$ particle are
equal to only the square root of $R_{pp}$ viz., $\lambda=r_q=s.$
Such choice so to speak 'kills two birds with one stone', as
follows. Let us recall 'if the creation of $pm$-Action ...
coincides with the range of past-present contact, ... $pm$-energy
is not range-decimated' (after Eq.\ref{gamma}). Hence if $q$
particles happen to be arranged sequentially at mutual distance
$r_q=\lambda$ from each other, the $c^2$ $pm$-energy propagating
through the sequence will be conserved. This constitutes the
'first bird'- effective causal contact over distance $r_q$ which
is $s$ times longer than $r_p.$ How to contrive such a sequence.
The planck particle defining Uncertainty $\hbar c=r_p(m_pc^2)$ may
be rewritten as $\hbar c=r_q(c^2 m_p/s)$ showing that the mass of
a $q$ particle denoted $m_q$ is equal to $m_p/s$ and therefore
that one planck particle generates $s$ such particles. This
constitutes the 'second bird'. If $s\times q$ particles are
generated in a sequence, its length $s\times r_q=s$ bridges the
inter-$p_+$ distance $R_{pp}=s^2$ as needed. However two problems
arise: how to contrive the 'lucky' generation of $q$'s along the
straight line of $\lambda$'s and how to ensure contiguous
connection of these lines (strings) in spacetime.

 {\it Low energy channel creating $c^2m_q$ cohesive energy
per each $\lambda-q$ couple.} An additional assumption becomes
necessary: NVF generates $s\times q$ particles in a stepwise
manner. In that case the 'lucky' generation of $q$'s along the
line of $\lambda$'s is explained as follows. A new generation is
started by an initial $\lambda$ having the $c^2$ $pm$-energy,
which has been emitted by an 'old' generation at the point of
vanishing. (As we shall see such immediate replacement is implied
by contiguous connectedness). A first $q$ is generated precisely
in the path of the initial $\lambda$ because the resultant
$\lambda-q$ couple creates a first section of a {\it channel}
having a $c^2m_q$ cohesive energy. The mass enables further
propagation of the signal and a second $\lambda$ continues the
first. In this manner the growing channel enables the generation
of a sequence of $q$'s, enabling the propagation of $\lambda$'s
from one mass to another, enabling the channel to grow, etc. As we
shall see a channel bifurcates occasionally forming a {\it
network} of interconnected straight channels/branches randomly
oriented with respect to each other. The network is named
$'\kappa$-Cluster'. In $s$ steps the target summed mass $sm_q=m_p$
is attained  and the particular generation of $s\times q$ ceases.
A $\lambda$ having the $c^2$ $pm$ energy is re-emitted from the
particular branch in which the target mass $m_p=s\times m_q$ has
been attained. This brings us back to a new beginning.

   {\it Contiguously connected $\kappa$-clusters
having $s^{1/2}$ branches of length $R_{\kappa}=s^{1/2}r_q$ each.}
Let us denote the lifetime (generation-time) and radius of the
$\kappa$-cluster by $t_{\kappa}$ and $R_{\kappa}$ respectively.
Since the $\lambda$'s propagate in straight line, $R_{\kappa}$ is
equal to $t_{\kappa}.$ The volume of a $\kappa$-cluster and the
volume per one excess $p_+$ are $R_{\kappa}^3$ and $R_{pp}{^3}$
respectively (to recall $R_{pp}=(r^3/r)^{1/3}=s^2).$ One
$\kappa$-cluster is generated from one excess $p_+$ and one excess
$p_+$ is created in the $R_{pp}{^3}$ volume during unit present
time. However the generation of each  $\kappa$-cluster requires
$t_{\kappa}$ units of time which define the time scale relevant to
what follows. During this 'extended present time', $t_{\kappa}$
excess $p_+$ and therefore $t_{\kappa}$-clusters and
$t_{\kappa}\times R_{\kappa}^3$ volumes, are generated inside the
$R_{pp}{^3}$ volume, getting us a {\it third} bird. Let
$\Phi_{\kappa,pp}$ denote the volume fraction of $\kappa$-clusters
in the $R_{pp}{^3}$ volume. Focal assumption is that the
$t_{\kappa}\times R_{\kappa}^3$ volumes fill the $R_{pp}{^3}$ and
ipso facto the entire $r^3$ causal volume contiguously, leaving no
Voids. Inter-cluster voids would imply never observed large scale
fluctuation of $\Gamma$ and (in terms of the modified Mach
principle), of rest mass of matter. I therefore require
$\Phi_{\kappa,pp}=t_{\kappa}R_{\kappa}^3/R_{pp}{^3}=1$ which gives
$R_{\kappa}=s^{3/2}.$ This implies that typical length of the
straight-line branches is also on the order of $s^{3/2}$ (or less
if many sequential branches add up to $R_{\kappa};$ however
subsequently a 'Growth model' supports branch length=$s^{3/2}$).
It consists of $q$ particles of length $r_q=s.$ Therefore one
branch/channel constitutes a linear sequence of $R_{\kappa}/r_q$
viz., of $s^{1/2}$ $q$-particles. However a $\kappa$-cluster is
generated from one $m_p$ mass, viz, it consists of $s\times
q$-particles. We conclude that the cluster (on the average)
constitutes a $d3$ network of $s^{1/2}\times R_{\kappa}$-branches,
consisting of $s^{1/2}$ $q$-particles each. We note further that
our result that the $R_{\kappa}^3$-volumes fill the $R_{pp}{^3}$
volume without voids, implies an equality of their respective mass
energy densities. Since the density inside the $R_{pp}{^3}$ volume
is equal to twice the critical density (Eq.\ref{omega}), we get
$\rho_{\kappa}= \rho_{pp}=2\rho_c.$ Summarizing,
\begin{equation}
\label{given} R_{\kappa}=s^{1/2}r_q=s^{3/2}~~~~giving~~
network=s^{1/2}\times(R_{\kappa}-branches)~~and~~\rho_{\kappa}=
\rho_{pp}=2\rho_c .
\end{equation}
How to interpret the structure of the network and  especially why
should we have bifurcation. Another problem arises in connection
to our no-voids result, whose derivation is based on the 'extended
present time' $t_{\kappa}=s^{1/2}r_q$. Since the lifetime of a $q$
particle lasts only one $r_q,$ our derivation mixes vanished-past
and existing-now $q$'s  That of course is perfectly valid when
dealing with a causal effect of past signals upon present time
particles. However Eq.(\ref{given}) describes a present time
'no-voids' behavior of $\kappa$-clusters and nonetheless a branch
of length $R_{\kappa}=s^{1/2}r_q$ consists of $s^{1/2}-1$ past $q$
particles and only a single existing one. I offer the following
reply. In order to decide presence or absence of a rapidly
generated-annihilated $\kappa$-cluster inside an $R_{\kappa}^{^3}$
volume, our experiment has to be equal to the generation time
$t_{\kappa}.$ We realize that our 'extended present time' does not
represent an ad hoc assumption (that gets us the third bird) but a
restriction in the spirit of the Uncertainty principle.

   {\it Termination-bifurcation growth}
In order to explain the bifurcation and network formation I assume
that starting from an origin the $\kappa$-cluster 'Grows' in
connected time steps. The steps do not grow in a single linear
sequence but constitute instead a 'tree' of simultaneously growing
sequences of steps or 'branches' each of which represents a
channel. At a time $t'$ we have a set of still growing branches.
Each of them may continue to grow linearly, or terminate or
bifurcate, increasing the number of $q$-particles viz. $s',$ by
one, zero or two respectively.  The varying outcomes I attribute
to random fluctuation of the $q$-generating process. The linear
increase by one merely gives a constant factor and may be ignored.
Termination and bifurcation must have precisely equal probability
such that non-exponential growth may continue indefinitely.
Despite this precise equality, fluctuation occurs and (in the
fashion of a random walk) an $rms$ resultant causes net
bifurcation. Termination at branch's end implies that it ceases to
grow permanently; if all branches terminate growth of chain stops.
Bifurcation implies that two concurrently existing $q$'s added at
branch's end encounter the excluded volume restriction. Hence I
assume that the branch/sequence of steps bifurcates, one channel
continues its straight line and the other veers off in a random
direction and subsequently continues a new straight line. The
subject has been studied in connection to the correspondence of
the percolating cluster to branched polymer; our case corresponds
to a 'T' (termination limited) Growth Model of clusters and of
polymers\cite{zevi}, with the following adaptations to our
$\kappa$-cluster: Both the number and length $r_q$ of each step
are equal to $s;$ there is no excluded volume (except between the
concurrently existing neighbor $q$'s that cause bifurcation) and
our channels  grow in straight line. Thus adapted T-Growth is
equivalent in mean field approximation, to a cluster consisting on
the average of $s'^{1/2}$ branches of length $r_q s''^{1/2}$ each.
However the product of instant $s'^{1/2}$ and $s''^{1/2}$ values
must reproduces the $sm_q=m_p$ mass precisely. We have seen that
what counts in the context of balanced expansion is the generation
of the $c^2m_p$ cohesive energy in each $\kappa$-cluster in
separate. The causal sphere constitutes therefore an extensive
system of $\kappa$-clusters (Eq.\ref{given}). Finally due to their
contiguous connection, an $c^2$ signal emitted by a vanishing
$\kappa$ cluster is immediately intercepted by a newborn $q$
particle, initiating the growth of a new $\kappa$ cluster.

\noindent{\bf  $5$ CREATION OF NUCLEONS AND RADIATION AT $0.12Gyr$
on $r^{1/3}$ and $r^{1/2}$ SCALES.}

 {\it Brief preview:} Let me propose the following description
of an instantaneous transition which had occurred at a
surprisingly recent radius time $r_*=0.7\times 10^{59}\sim
0.12Gyr.$ All $(\bar q-\tilde q)_+$ pairs of which consists DE
converted on the $s_*$ scale in equivalent parts as follows: the
$\bar q_+$ into nucleons $(n)$ and the $\tilde q_+$ into high
energy $\gamma$ photons. The energy of each nucleon and each
photon corresponded to $c^2m_p/s_{*}\sim.9GeV.$ The transition
occurred in all $\theta_2 r_*$ clusters, whose growth had started
precisely at time $r_*,$ terminated at $r_*+t_{\kappa}$ ($0.12Gyr$
to $0.12Gyr+10^{-14}s$) and which kept their $s_*$ constant.
Thermal equilibrium and Baryogenesis obtained inside
$\kappa$-clusters which constituted disjoint islands isolated from
each other, at an energy density of nucleons and photons of about
$c^2m_p/s{_*}^{9/2}\sim.7MeV,$ larger than critical by
$2r{_*}^{1/2}.$ Subsequent very rapid 'global' dispersion of the
$\kappa$-clusters over the causal sphere turned the $\gamma$
photons into starting CMB and the $n$'s into starting mix of
nucleons ('cosmic baryons'). Their respective fractions on global
scale were $\Omega_{*,cmb}=\Omega_{*,n}=\theta_2$ (at the expense
of $\Omega_{*,de}).$  The photons decoupled from matter and
attained equilibrium at a 'warmish' $T_{*,cmb}=326K.$ From $\simeq
r_*$ and until now CMB redshifted due to cosmic expansion by
$r_0/r_*\simeq 115,$ to $\simeq 2.83K,$ versus accepted $2.73K$
(fair agreement considering that it is derived from the proton
mass and Hubble's $H_0$ alone).  The evolution of nucleons was
more complex. Since their creation had ceased while that of DE and
DM continued as usual, to a first approximation $\Omega_{n}$ too
decreases by $r_0/r_*.$ However the absence of continuing creation
promotes the clustering of $n$'s. I therefore propose a model of a
Global nucleonic cluster which constitutes a self similar copy of
the $\kappa$-cluster, scaled-up from radius $R_{\kappa}$ to $r.$
The cluster's $\Omega_{n}$ decreases more slowly than predicted by
first approximation namely only by $(r_0/r_*)^{2/3}.$ The model
fits today's $\Omega_{n,0}$ and also seems to explain the
extraordinary dimming of far away Super Novae. The latter is
commonly attributed to Accelerated Expansion but here it is
attributed  to nucleonic clustering.

{\it In detail:} According to Eq.(\ref{de}) but in terms of $q$'s,
each DE cluster consists half by half of virtual $\bar q_+$ and
$\tilde q_+.$  The separation distance between these particles
(irrespective of their type) is $s$ and is bridged contiguously by
the wavelength $\lambda=s.$ However in the context of their
conversion into nucleons, the $\bar q_+$ skipping over $\tilde
q_+$ became interconnected by twice longer wavelength
$\lambda_{2s}=2s,$ which bridged the distance between
second-nearest neighbors of the same type. Specifically a newly
created real $n$ particle emitted a $\lambda_{2s}$ which formed
the low energy channel favoring the conversion of its
second-nearest neighbor from $\bar q_+$ to $n.$ The channel could
not link the new $n$ to a $\tilde q_+$ because that would result
in gratuitous energy being released by annihilation. Each
$\lambda_{2s}$-$\bar q_+$ couple generated one $n.$ The associated
energy was $e'_{2s}=2\pi\bar hc/2r_q=\pi c^2m_p/s$ (substituting
$\bar hc/r_p$ by $c^2m_p).$ At the point of transition, $e_{2s}'$
was precisely equal to a nucleon's rest energy $c^2 m_n$ giving
\begin{equation}
\label{recent} r_*\equiv s_{*}^3=(\pi m_p/m_n)^3=0.069\times
10^{60}~(\propto 0.12Gyr/t_p~~or~~z\sim 114).
\end{equation}
In parallel, each deserted  $\tilde q_+$  antiparticle attained by
a $\lambda$ converted into a (real) high energy $\gamma$ photon.
No $\lambda_{2s}$ was involved because no causal contact with past
particles was needed. The creation of an $n$ with the help of
twice as long $\lambda_{2s}$ involved an energy one half as large
as the  energy involved in the creation of a $\gamma.$ The half as
large energy per step implies that twice as many, twice as long
steps were required in order to create nucleons whose total rest
energy was equal to that of $\gamma$ photons. I conjecture that
this disparity led to a 'kernel' cluster of $\gamma$ and of some
$n$'s surrounded by a 'halo' cluster consisting of remnant $n$'s
alone. Let us denote the radii of the kernel and the halo clusters
by $R_{\gamma n,kern}$ and $R_{n,halo}$ respectively. Clearly
$R_{\gamma n,kern}=R_{\kappa}$ because both were created with the
help of the $\lambda$ steps. The conversion at $r_*$ was creating
the $\gamma$'s $2^{3/2}$ times faster than the $n$'s. After time
$t_{\gamma n,kern}(=R_{\gamma n,kern})$ it had completed the
creation of a the kernel cluster comprising all $\gamma$ and only
a fraction of $n$. Thereafter the creation of $n$'s continued till
time $t_{n,halo}=2^{3/2}t_{\gamma n,kern}$ creating a 'halo'
cluster of remnant $n$'s, initially devoid of $\gamma$ photons and
having radius $2^{3/2}$ larger than the kernel cluster. The
corresponding mass energy densities were
$\rho_{n,halo}=2^{-(3/2)x}\rho_{\gamma n,kern}$, where $x$ will be
determined at once. Let us compare both to a contemporary density
of unconverted $\kappa$-clusters of Eq.(\ref{given}). To recall
the number of $\kappa$-clusters in the $R_{pp}{^3}$ volume is
$t_{\kappa}$  and they fill the $R_{pp}{^3}$ volume contiguously.
However at anytime they contain only $s^{1/2}$ existing 'now'
$q$-particles, the rest belongs to vanished 'past' $q$-particles.
Momentarily let us ignore the $2\pi$ and $\theta_2$ factors and
the relatively minor disparity between the $\gamma n,kern$ and the
$n,halo$ clusters. Hence referring only to the creation of the
$n,halo$ we note that (in contrast to $\kappa$-clusters) all
$q_+$'s converted into real $n$ leaving no vanished past
$q$-particles behind. Hence the number of $n$'s contained in the
$R_{n,halo}{^3}$ volume was $s.$ These $s\times m_n$ masses
represent the entire $m_p$ mass contained in the $R_{pp}{^3}$
volume. Conclusion the energy density of the single $n,halo$
cluster contained in  $R_{pp}{^3}$ was larger by
$(R_{pp}/R_{n,halo}){^3}=r^{1/2}$ than that of $\kappa$-clusters
but in compensation its volume fraction was correspondingly
smaller, $\Phi_{n,halo}=(R_{pp}/R_{n,halo})^{-3}=r^{-1/2}$ and
neighbor $n$ clusters in causal sphere became disjoint. Auxiliary
conclusion: $x=3.$ Thereafter the $n,halo$ and the $\gamma n,kern$
had rapidly spread over the $R_{pp}{^3}$ volume (hence over entire
causal sphere and practically still at time $r_*$), their density
dropped to $\rho_{n;\gamma,pp}=(\theta_2/2)\rho_{\kappa}=
\theta_2\rho_{c,*}.$ (The $\theta_2/2$ factor is due to equal
sharing between $n$ and $\gamma;$ I surmise that the $2\pi$ factor
drops out upon averaging the $e'_{2s}$ energy of $\gamma_{2s}$
wavelengths (Eq.\ref{recent}), over the $\kappa$ cluster). We get
\begin{equation}
\label{kappa-n} R_{\gamma
n,kern}=2^{-3/2}R_{n,halo}=R_{\kappa};~~~ \rho_{\gamma
n,kern}=2^{9/2}\rho_{n,halo}=2r_{*}^{1/2}\rho_{c,r_*}~and~~
\rho_{n;\gamma,pp}=\theta_2\rho_{c,r_*}.
\end{equation}
Following the transition and Baryogenesis inside disjoint
$\kappa,n$ clusters, the latter were cooling due to escape of
photons and global dispersion over entire causal sphere. The
effect of this global dispersion corresponds in Eq.(\ref{kappa-n})
to the passage from $\rho_{\gamma n,kern}=
2r_{*}^{1/2}\rho_{c,r_*}$ to
$\rho_{n;\gamma,pp}\simeq\theta_2\rho_{c,r_*},$ giving an
immediate estimate of a decrease of temperature by $s_*{^{3/8}}.$
More explicitly $\Omega_{*,n}=\Omega_{*,\gamma}=\theta_2.$
Radiation Law combined with $\rho_{*,c}=1.3\times
10^{-25}gcm^{-3}$ (Eqs.\ref{omega} and \ref{recent}), gives the
CMB starting temperature $T_{*,cmb}=326K.$ Thereafter the
evolutions of nucleons and of CMB diverge. Let us define
$\alpha=r/r_*.$ While $\alpha$ increased from one to today's
$\alpha_0,$ CMB photons redshifted in uniformly expanding space,
giving $T_0=T_{*,cmb}/\alpha_0.$ Hubble's $h_0=0.72$ gives
$\alpha_0=115.$ Thus at $r\simeq r_*$ and at  $r_0$ we have
\begin{equation}
\label{*} \Omega_{*,cmb}=\Omega_{*,n}=\theta_2;~~ T_{*,cmb}=326K
~and~{T_{0,cmb}=T_{*,cmb}}/\alpha_0=2.83K.
\end{equation}

{\it Nucleonic evolution at $1<\alpha\le\alpha_0.$} The transition
created within the causal sphere a total of $r_*s_{*}=s{_*}^{4}$
nucleons of mass $m_n$ each, a total conserved at all $\alpha$
(with neglect of creation of higher atomic masses, luminous matter
and temporarily of $\theta_2,$ and of $2^{9/2}$). I stipulate that
after dispersion of nucleons on the global $r_*$ scale they formed
a Global nucleonic $(Gn)$-Cluster. Its structure is determined by
that at all $\alpha\ge 1,$ the distance separating neighbor
nucleons was and to this day is equal to the current value of
$\lambda=r_q=s$ separating $q$ neighbors in a $\kappa$-cluster
(Eq.\ref{given}). This enables the $c^2$ $pm$-energy to propagate
in a $Gn$-cluster like everywhere else in space, with the help of
$\lambda$'s in the low energy channels. As we shall see this
equivalence of the $n-n$ and $q-q$ distances implies that the $Gn$
cluster expands slower than expanding space. Consequently the low
energy generated by the $Gn$ has to be devoted to 'internal
needs', namely to opposing the pull of expanding space trying to
stretch the $Gn$ structure. It therefore seems to me that although
nucleonic matter raises total $\Omega$ above one, its extra
cohesive energy is devoted to aforesaid internal needs and does
not perturb uniform expansion of space. I also assume that at
creation time $(\alpha=1)$ the $Gn$ cluster consisted of $s{_*}^2$
straight branches, of radius $R_{*,Gn}=s{_*}^2\lambda_*=r_*$ each.
The assumption amounts to mere scaling-up of the 'TB'
Model\cite{zevi} of a $\kappa$-cluster to the $Gn$-cluster: The
former consists of $s$ particles distributed into $s^{1/2}$
branches of radius $s^{1/2}\lambda$ each.  The latter consists of
$s{_*}^4$ particles distributed into $s{_*}^2$ branches of radius
$s^2\lambda_*$ each. Thereafter while the causal radius expands by
$\alpha,$ the network structure is essentially conserved. The only
variation is a stretching of the $\lambda$ component of $R_{*,Gn}$
which (as has been stipulated) expands from $\lambda_*=s_*$ to
$\lambda_{\alpha}=\alpha^{1/3}s_*.$ Consequently
$R_{\alpha,Gn}=\alpha^{1/3}R_{*,Gn}=\alpha^{1/3}r_*.$ The $Gn$
cluster network of channels contracts and glides effortlessly
inside expanding causal sphere, because the cohesive energy of
channels is entirely devoted to opposing the restricted expansion
of its own radius beyond the allowed
$R_{\alpha,Gn}=\alpha^{1/3}R_{*,Gn}$ increase. At $\alpha=1,$
$\Omega_{*,n}$ is critical (see Eq.(\ref{*} with neglect of
$\theta_2),$ implying $\Omega_{*,n}\propto R_{*,Gn}^{-2}$
(Eq.(\ref{omega}). Hence $\Omega_{\alpha,n}$ decreases due to the
expansion of $R_{\alpha,Gn}^{-2}$ as $\alpha^{-2/3}.$ The result
refers to conditional $\Omega_{\alpha,n}$ (or radius
$R_{\alpha,Gn}),$ stipulating that both the event and the observer
belong to the $Gn$ cluster. If one of them belongs to a 'void',
the corresponding value will tend to the global average
$(\Omega_{\alpha,n})_{glob}\propto\alpha^{-1}.$ This may be
quantified by measurements of the distance from us to an Event',
where both 'us' and the event belong to the $Gn$ cluster. Suppose
two distances $R_0$ and $R$ are measures at two respective times
$\alpha_0$ and $\alpha.$ The $R_0-R$ difference and the
$\alpha_0-\alpha$ one are determined respectively by the expansion
of the $Gn$ cluster $\propto\alpha^{1/3}$ and of the causal sphere
$\propto\alpha.$ We get $(R_0-R)/(\alpha_0-\alpha)$ equal to
$(\alpha/\alpha_0)^{2/3}$ and the latter is equal to
$[(t_0-\ell/c)/t_0]^{2/3}=(1-\ell/\ell_H)^{2/3}$ where $\ell$ is
the 'light'-distance corresponding to $\alpha_0-\alpha$ and
$\ell_H$ is the 'Hubble-length' (viz. the causal radius). In
summary
\begin{equation}
\label{now}{R_{\alpha,Gn}\over r}=\alpha^{-2/3};~~~~
({\Omega_{0,n}\over\Omega_{*,n}})_{_{_{Gn}}}=\alpha_0^{-2/3}=0.042;~~~~
{(R_0-R)_{Gn}\over\alpha_0-\alpha}=(1-\ell/\ell_H)^{2/3}.
\end{equation}
Substituting $\Omega_{*,n}=\theta_2$ we get $\Omega_{0,n}=0.029,$
lower than, but not excluded by most data\cite{book}. In contrast
the first approximation gives $\Omega_{0,n}=0.006$ which
definitely seems too low, except when referring to a global
average density. It should be also stressed that Eq.(\ref{now})
describes the decrease of $\Omega_{0,n}$ with time $\alpha_0;$
{\it it does not} describe nucleon concentration during
baryogenesis. A decrease of $\Omega_{\alpha,n}$ with $\alpha$ is
inherent to our model of constant creation of DE and DM as opposed
to the singular joint creation of nucleons and of CMB. However its
precise form given in Eq.(\ref{now}) involves a conjectured
existence of the $Gn$ cluster. The ratio
$R_{Gn}/r=\alpha^{-2/3}=0.042$ combined with experimental
$r\approx4100Mpc$ puts the radius of a 'supreme' cluster at
$R_{Gn}=170Mpc,$ judged to be in fair agreement with a reported
$R_{supr}\approx100Mpc.$ The result strongly suggests that our
causal sphere contains huge voids which are indeed observed. The
fractal structure of the $Gn$ cluster proposed here may be
examined with the help of data on inter-galactic distances in
clusters and super-clusters of galaxies. (Famous 'box' algorithm
seems not suited for the purpose, an algorithm tailored for that
seems to be 'growth of clusters'\cite{ibes}).

\noindent{\bf $6$ BARYOGENESIS; DECELERATION; DIRAC NUMBERS;
CONCLUSIONS.}

{\it Baryogenesis in $\kappa$-clusters.} Baryogenesis occurred in
the $\gamma n,kern$ cluster right after its creation at $r_*.$
First using $\rho_{\gamma n,kern}$ of Eq.(\ref{kappa-n}) and the
radiation law gives $T_{\gamma n,kern}=0.69MeV$ which matches very
well the value associated with the BB baryogenesis (BBN) namely
$T_{bbn}=0.7MeV$\cite{book}. Second, using $\rho_{n,halo}$ and
$\rho_{\gamma n,kern}$ of Eq.(\ref{kappa-n}) we find that a
'nucleon concentration' defined as the ratio of the two energy
densities equals $2^{-9/2}=0.044.$ This concentration lies within
bounds derived from the abundance of light elements\cite{book}. We
have used the ratio of $\rho_{n,halo}$ to $\rho_{\gamma n,kern}$
because in our model the former has determined the concentration
of nucleons and the latter has determined the $T_{\gamma n,kern}$
temperature that existed during baryogenesis. The same result
obtains if we base the comparison on a 'critical density' invoked
in the literature\cite{book} of an 'BB-equivalent' causal sphere
having a radius we denote $r_{bbeq}$ as follows. Our model of
baryogenesis had occurred in a cluster of radius  $R_{\gamma
n,kern}$ of Eq.(\ref{kappa-n}), enclosing one $m_p$ mass. A
corresponding radius enclosing $rm_p$ masses would be enclosed by
$R_{\gamma n,kern}{^{3/2}}.$ Hence we expect $r_{bbeq}=R{_{\gamma
n,kern}}^{3/2}.$ Critical energy density is inversely proportional
to radius squared (Eq.\ref{omega})., hence $\rho_{c,bbeq}\propto
r_{bbeq}^{-2}\propto R_{\gamma n,kern}^{-3},$ giving again the
$2^{-9/2}$ nucleon fraction but more in the manner of BBN. Third,
the BB-equivalent radius $r_{bbeq}$ also gives $t_{bbeq}=35s$
approximating $t_{bb}\simeq 100sec$\cite{book}. The three
similarities  give rise to hope that a description based on our
model will agree with well established parameters of baryogenesis.

 {\it (pseudo) 'Acceleration'=decelerated expansion of the $Gn$ cluster}
In Eq.(\ref{now}) we have described how the recession of a pair of
nucleon 'particles' from each other (be it intergalactic dust, or
'us' and a SNe $1$a) is decelerated due to that both belong to the
$Gn$ cluster. The deceleration affects the recession as measured
by the redshift $z$ of the flash of light reaching from an SNe
$1a$ but not the spacetime 'luminosity distance' $\ell$ travelled
by this flash with speed $c$ (measured for example by the decrease
of apparent brightness). We identify the redshift-measured
distance $z$ and the luminosity-measured distance $\ell$
respectively with $R_0-R$ and with $\alpha_0-\alpha$ of
Eq.(\ref{now}) and get
\begin{equation}
\label{acceler}
z/\ell=(1-\ell/\ell_H)^{2/3},~~~z,~\ell~and~\ell_H~=~
redshift,~luminosity~and~Hubble~distances.
\end{equation}
In the range of available data the result resembles $\ell/z\approx
1+z$\cite{filippenko}. It seems to me that the great scatter of
data testifies to the fact that certain SNe $1a$ are separated
from us by a substantial void(s). Possibly the scatter reveals an
anisotropy which may enable us to sort the network of our super
cluster of nucleonic matter from self similar voids on all scales.

{\it Dirac numbers related to inter-planck distance.} An enigma
considered by Dirac\cite{dirac} is that seemingly unrelated
dimensionless numbers are on the same large order of $\approx
10^{40}.$ I claim that Dirac numbers represent the $s^2\simeq
10^{40}$ reduced distance between excess planck particles created
at present or more precisely during recent proton-creating
transition. The numbers are derived from ratios involving
$r_0\simeq 8.8\times 10^{60},$ $s_*/\pi(=m_p/m_n)=1.3\times
10^{19}$ (Eq.(\ref{recent}) and $s_{*}^2.$ On this basis the
following numbers are reproduced precisely: Ratio of electrostatic
and gravitational interaction of proton-electron pair $N_1\equiv
e^2/Gm_p m_e$ becomes $N_1=(1800/137)(s_*/\pi)^2.$ Ratio of the
proton Compton and Schwarzschildt radii $N_2\equiv (\hbar c/m_n
c^2)(Gm_n/c^2)^{-1}$ becomes $N_2=(s_*/\pi)^2$. Ratio of $r_0$ to
classic electron radius $N_3\equiv r_0(e^2/m_e c^2)^{-1}$ becomes
$N_3=(137/1800)r_0/(s_*/\pi).$ Present time number of protons
whose density is taken to be critical is $N_4\equiv(4\pi
r_0^3/3m_n)\rho_{0,c}$ becomes $N_4=r_0/(s_*/2\pi).$

  {\it General remark.} Admittedly a shortcoming of the present
model is an absence of unified mathematical formulation. Still
many of its aspects are more amenable to study with the help of
simulation, like the shape of the $\kappa,n$ and especially of the
$Gn$ cluster. Another remark is that possibly right from start the
model should be formulated systematically in terms of
fluctuations. One well known outcome of fluctuation is to bring
about critical behavior: viz., oppositely acting and equally
probable fluctuations generate a vastly smaller but non-vanishing
$rms$ resultant. Here $r^4$ fluctuations which with equal
probability either add or subtract, either planck particles or
antiparticles generate for example $r=((r^4)^{1/2}){^{1/2}}$ real
planck particles. However a different but equally important role
of fluctuation combined with Uncertainty is to bring into causal
contact particles which appear to be isolated from each other {\it
without invoking Action at Distance}. A case in point is the set
of $r$ excess $p_+$ separated by the seemingly insurmountable
distance $s^2=(r^3/r)^{1/3}.$  Assistant conditions are needed.
First, assumedly each $m_p$ mass may be generated not at one go
but instead 'Grow' in a sequence of steps producing each a subunit
virtual $q$ particle. The particles are $s$ times lighter hence
more numerous but also (by Uncertainty) $s$ times larger. This
already enables to link the excess  $p_+$'s with the help of
straight lines (strings) consisting of a sequence of $q$ enabling
a creation/propagation of the cohesive energy that balances the
expansion. However the straight lines leave out vast voids
unconnected. A model of a branching-out chain describing
Percolation helps us to device a model of a cluster of $q$
particles which fills space contiguously. In conclusion,
randomness, fluctuation, clusters and probability seem intimately
linked to spacetime.

  {\it Tests of the Model:} Fundamental tests are uniform expansion,
constantly created critical mass energy density $\Omega=1$ and
generation of Dark Matter and Dark Energy, constantly obeying
$\Omega_{dm}\simeq0.3$ and $\Omega_{de}\simeq0.7.$ Passing these
tests is mandatory but somewhat inconclusive because they fit well
known results and as such are not really predictive. However our
model of nucleonic matter bound together into a Global Cluster on
the scale of the causal radius does predict that $\Omega_n$ should
decrease with time as in Eq.(\ref{now}); it also predicts that the
amazing Acceleration is actually attributable to the Global
Cluster of nucleons, as described by Eq.(\ref{acceler}) and as
such provides information not on cosmic expansion but on the
structure of (inevitably  declining) nucleonic fraction;
supportive in this context is our estimate of the radius of the
largest galactic super-cluster and explanation of great voids;
another prediction is that the well documented BBN theory may be
rewritten in terms of our model of a baryogenesis which had
occurred on the $r^{1/2}$ scale of $\kappa$-clusters as late as at
$0.1Gyr.$ Passing these tests would be supportive however in
separate they are not mandatory because each requires an extra
assumption (except for mandatory but qualitative prediction that
$\Omega_{n}$ was much larger in the past).


\bibliographystyle{srt}


\end{document}